\documentclass{PoS}
\usepackage{wrapfig}
\usepackage{amsmath}
\usepackage{amssymb}
\usepackage{graphicx}
\usepackage{epsfig}
\usepackage{bm}

\title{Scale evolution of kt-distributions}

\ShortTitle{Scale evolution of kt-distributions}

\author{\speaker{Federico Alberto Ceccopieri}\\
	Vrije Universiteit Brussels, \\
        Interfaculty Institute for High Energy (IIHE)\\ 
        Pleinlaan 2, 1040 Brussels,  Belgium \\
        E-mail: \email{federico.alberto.ceccopieri@cern.ch}}

\abstract{We review some of the features of the evolutions equations 
for transverse momentum dependent parton distributions recently proposed by us.
We briefly describe the new ingredients entering the equations and 
their relationship with ordinary evolution equations.
We comment on possible choices for the initial conditions 
and then show results for the evolved distributions 
obtained by numerical implementation of the equations. 
By computing the average transverse momentum at different scale 
we highlight some general properties of the solutions.}

\FullConference{XVIII International Workshop on Deep-Inelastic Scattering and Related Subjects\\
                 April 19 -23, 2010\\
                 Convitto della Calza, Firenze, Italy}
                                            
\begin{document}

\section{Introduction}
	
Transverse momentum dependent, or equivalently, $k_t$-distributions 
are currently object of intense research activity due to 
their wide range of applicability in the description of hadron initiated 
hard processes~\cite{Hautmann_Jung,WMR}.
The use of $k_t$-distributions is indeed phenomenologically appealing since observables 
constructed upon them show a reasonable agreement with data
already in lowest order, which is not the case for predictions based  
on collinear factorization at the same accuracy, see for instance the 
discussion in Ref.~\cite{Zpt_Kwiecinski}.
Their correct formalization in quantum-chromodynamics appears to be well established only
at high energy~\cite{kt_fact}. Away from this limit the situation is still unclear, 
although detailed investigations are present in the literature~\cite{CS,Stefanis,Hautmann}. 
The relevant issue of factorization within a $k_t$-dependent approach has also been investigated~\cite{Ji}.
The latter property together with a precise knowledge of the scale dependence of $k_t$-distributions  
would allow to relate to each other data coming from 
experiments at different energies and, more important, to test factorization
quantitatively.
Although a definitive answer is absent in the literature, 
there have been however some attempts, see for example~\cite{Kwiecinski1loop,Kimber}. 
In particular we will focus on the equations proposed by us 
in Ref.~\cite{CT} which are the space-like version of the ones proposed 
in Ref.~\cite{BCM}. 
In a subsequent phemonelogical study~\cite{CT2}, performed in 
the context of semi-inclusive deep inelastic scattering, 
it was shown that a reasonable description of data could be 
obtained once unintegrated evolution equations 
were solved with suitable, but motivated, initial conditions 
and assuming factorization for the cross-sections of interest. 
This result stimulated us to apply the same formalism to the description 
Drell-Yan type processes in hadronic collisions~\cite{Zpt_Ceccopieri}.
The $p_t$-spectrum of the gauge boson has, in fact, a rich 
transverse structure and involves both perturbative and 
non-perturbative aspects of the underlying theory.
The resummation of the perturbative series in the multiple soft gluon emission limit 
can be accomodated by using properly modified unintegrated evolution equations~\cite{KT} 
so that the structure of the non-perturbative form factor can be investigated.

\section{Space-like $k_t$-dependent evolution equations}	
\begin{figure}[t]
\begin{center}
\includegraphics[width=2.5in]{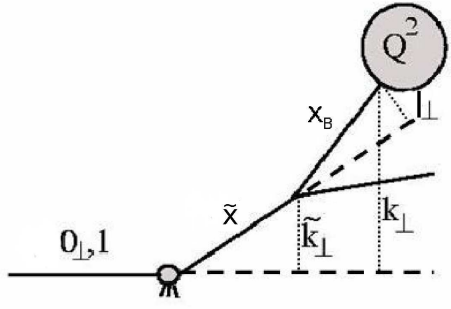}
\caption{Pictorial representation of the proposed evolution equation.
The longitudinal and transverse component of the partons involved in the splitting 
$\widetilde{k} \rightarrow k +l$ are indicated. \protect\label{fig1}}
\end{center}
\end{figure}

Transverse momentum dependent parton distribution function, 
$\mathcal{F}_{P}^{i}(x_B,Q^2,\bm{k_{\perp}})$, give
the probability to find, at a given scale $Q^2$, a parton $i$ 
with longitudinal momentum fraction $x_B$ and transverse momentum $\bm{k}_{\perp}$ 
relative to the parent hadron momentum.
The evolution equations for $k_t$-distributions proposed in Refs.~\cite{CT,CT2} read
\begin{multline}
\label{dglap_TMD_space}
Q^2 \frac{d \mathcal{F}_{P}^{i}(x_B,Q^2,\bm{k_{\perp}})}{d Q^2}
=\frac{\alpha_s(Q^2)}{2\pi}\int_{x_B}^1 \frac{du}{u^3} 
P_{ji}(u) \cdot\\ \cdot \int \frac{d^2 \bm{l}_{\perp}}{\pi}\,\delta(\,(1-u)Q^2-l^2_{\perp})
\,\mathcal{F}_{P}^{j}\Big(\frac{x_B}{u},Q^2, \frac{\bm{k}_{\perp}-\bm{l}_{\perp}}{u} \Big)\,,
\end{multline}
and resum large logarithms associated to the emission
of collinear partons. However, at variance with ordinary evolutions
equations~\cite{AP},  the transverse momentum generated 
at each branching, $\bm{l}_{\perp}$, is explicitely taken into accont kinematically 
and eventually adds to the non-perturbative one due to  
Fermi motion of the parton in the parent hadron.
If we now consider $\mathcal{F}$ 
to be the bare distribution of a quark in a quark, neglecting flavour indeces,   
we may insert  the source term 
$\mathcal{F}(x,Q^2,\bm{k_{\perp}})=\delta(1-x)\delta^{(2)}(\bm{k}_{\perp})$
into eq.~(\ref{dglap_TMD_space}) obtaining, up to $\mathcal{O}(\alpha_s)$, the following result
\begin{equation}
\label{one_loop}
\mathcal{F}(x,Q^2,\bm{k}_{\perp})=\delta(1-x)\delta^{(2)}
(\bm{k}_\perp)
+\frac{\alpha_s}{2\pi}\frac{1}{\pi}\frac{1}{k_{\perp}^2} P(x)\,,
\end{equation}
which indeed exposes the singularity associated to the emission of a collinear parton
weighted by splitting functions $P(x)$~\cite{AP}. 
In order to clarify the role of the new ingredients appearing 
in eq.~(\ref{dglap_TMD_space}), it proves useful to 
consider just one emission, see Fig.~(\ref{fig1}). 
The small blob at the bottom left of the figure then symbolizes the iteration of emissions 
in the parton ladder of whose 
only the last, $\widetilde{k} \rightarrow k +l$, is explicitely shown. 
Transverse momenta $\widetilde{\bm{k}}_{\perp}$ and $\bm{k}_{\perp}$ are defined 
with respect to the incoming hadron direction, while $\bm{l}_{\perp}$ is defined 
with respect to $\widetilde{k}$ direction.
The transverse momentum $\bm{\widetilde{k}}_{\perp}$ 
and longitudinal momentum $\widetilde{x}$ of the branching parton $\widetilde{k}$ 
are written as a function of the relative transverse momentum $\bm{l}_{\perp}$
and the fractional momentum $u$ appearing in the splitting.
They can be expressed as $\bm{\widetilde{k}}_{\perp}=(\bm{k}_{\perp}-\bm{l}_{\perp})/u$ 
and $\widetilde{x}=x/u$, respectively. The former is a result of the transverse boost 
from $\widetilde{k}$ to hadron direction~\cite{GSW}. Both can be found in 
the distribution under convolution in the right hand side of eq.~(\ref{dglap_TMD_space}).
The additional integration $d^2 \bm{l}_{\perp}$and the $\delta$-function 
 appearing in eq.~(\ref{dglap_TMD_space}) 
are associated with the phase space and mass-shell contraint of the emitted partons. 
The latter can be written as 
\begin{equation}
\label{mass}
l^2_{\perp}=-(1-u)k^2+u(1-u)\widetilde{k}^2\,,
\end{equation}
where the emission of massless partons has been assumed ($l^2=0$). 
As is well known, logarithmically enanched contributions arise 
when the interacting parton virtuality increases along the ladder. 
Therefore in the limit $\widetilde{k}^2 \ll k^2$ we may neglect $\widetilde{k}^2$ with respect
to $k^2$ and set $-k^2=Q^2$, obtaining 
$l^2_{\perp}=(1-u)Q^2$, which can be found in eq.~(\ref{dglap_TMD_space}).  
Furthermore, we assume $k_t$-distributions to fulfil the normalization condition:
\begin{equation}
\label{spacelike_norm}
\int d^2 \bm{k}_{\perp} \mathcal{F}_{P}^{i}(x_B,Q^2,\bm{k}_{\perp})=
f_{P}^{i}(x_B,Q^2)\,,
\end{equation}
where $f_{P}^{i}$ are ordinary parton distributions.
It is important to remark that integrating eq.~(\ref{dglap_TMD_space}) over $d^2 \bm{k}_{\perp}$
and by using eq.~(\ref{spacelike_norm}) we recover the ordinary 
evolution equations for $f_{P}^{i}$~\cite{AP} .
\begin{figure}[t]
\centerline{\psfig{file=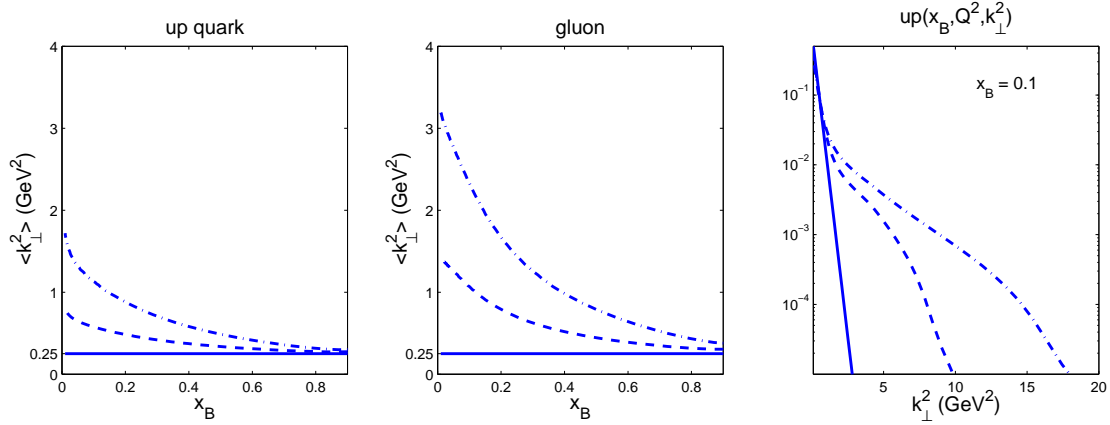,width=7.0in}}
\vspace*{8pt}
\caption{Left and middle panel: average transverse momentum $\langle k_{\perp}^2\rangle $ 
for up-quark and gluon at three different scales:
$Q_0^2=5$  Ge$V^2$ ($-$), $Q^2=10$ Ge$V^2$ ($--$) and $Q^2=20$ Ge$V^2$ ($-\cdot-$)\,.
Right panel: the transverse spectrum of the up quark distributions at fixed $x_B=0.1$ 
at the previously indicated three different scales.
\protect\label{fig2}}
\end{figure}

\section{Numerical solutions}
The unintegrated evolution equations, eq.~(\ref{dglap_TMD_space}), are numerically solved by 
a finite difference method on a discrete grid in $(x_B,\bm{k}_{\perp})$ space.
The initial conditions at the starting scale $Q_0^2= 5$ Ge$V^2$ 
are chosen to have the factorized form 
\begin{equation}
\label{ic}
\mathcal{F}_P^i(x_B,Q_0^2,\bm{k}_{\perp})=  f_P^i(x_B,Q_0^2) \frac{1}{\pi<k_{\perp,i}^2>}
\,e^{\frac{-k_{\perp}^2}{<k_{\perp,i}^2>}} \,\,\,
\;\;\; i=q,\bar{q},g
\end{equation}
where $f_P^i(x_B,Q_0^2)$ are ordinary parton distributions~\cite{GRV}.
The choise of a energy ($x_B$) independent guassian transverse factor
in eq.~(\ref{ic}) is supported, for example, by recent semi-inclusive deep inelastic scattering
data at low energy~\cite{Misha}. 
The width $\langle k_{\perp,i}^2\rangle $ are fixed to a value of $0.25$ Ge$V^2$
both for quarks and gluons~\cite{CT2}.
Since the initial conditions, eq.~(\ref{ic}), by construction fulfil eq.~(\ref{spacelike_norm}),
the latter can be checked for every values of $x_B$ and $Q^2$ to estimate and, eventually, to improve 
the numerical accuracy of the evolution. For this check to be meaningful we adopt 
the same flavour scheme and coupling values used in Ref.~\cite{GRV}.
The evolved distributions show indeed some interesting properties.
In the left and middle panel of Fig.(\ref{fig2}) 
we present the average transverse momentum, $\langle k_\perp^2 \rangle$, 
as a function of $x_B$ calculated at
three different scale for up quark and gluon, respectively. 
In both cases, the averaged transverse momentum at the final scale,  $Q^2= 20$ Ge$V^2$,
increases with decreasing $x_B$. The effect is more 
evident in the gluon case, due to the singular behaviour of $P_{gg}(u)$ at small $u$. 
In the opposite limit, $x_B \rightarrow 1$, the averaged 
transverse momentum approaches the value given in the initial condition and indicated by
the horizontal line in left and middle panel of Fig.(\ref{fig2}). 
In this limit, phase space only allows the emission of soft partons
which generate negligible trasverse momentum.
In the right panel of Fig.~(\ref{fig2}) it is shown the up-quark distribution at fixed $x_B=0.1$ 
as a function of $k_\perp^2$. 
Interestingly, the factorized form of eq.~(\ref{ic}) is not preserved under evolution. The guassian 
dependence on $k_\perp^2$ at $Q_0^2$ is turned into a inverse power law $1/(k_\perp^2)^a$
at the final scale $Q^2$, the latter dependence in qualitative agreement with matrix 
elements behaviour at large $k_\perp^2$, eq.~(\ref{one_loop}).
We also observe that, as $Q^2$ increases, the distributions start to \textit{populate}
the higher $k_\perp^2$ region, as an effect of the leading logarithmic approximation 
built-in the evolution equations.   
\section*{Conclusions}
In this contribution we have reviewed some of the features  
of the unintegrated evolution equations  proposed by us and discussed 
some peculiar properties of the evolved $k_t$-distributions obtained by numerical solution. 
The use of unintegrated distributions is indeed phenomenologically appealing~\cite{CT2,Zpt_Ceccopieri}.
Most interestingly, the analysis of jet observables within this framework may give access
to valuable informations on the unintegrated gluon distributions
which describes hadron structure as probed in high energy collisions. 
However, the departure from a pure collinear scheme  
introduces a number of non trivial problems which are at present 
under investigation by many groups. This implies that, to date, any conclusive statement 
on the validity of $k_t$-distributions approach in general, 
and on the proposed evolution equations in particular, must wait for further theoretical developments.

\end{document}